\documentclass[10pt,aps,prl,twocolumn,superscriptaddress,reprint]{revtex4-1}

\usepackage{lmodern}
\usepackage[version=3]{mhchem}
\usepackage{graphicx}
\usepackage{hyperref}
\usepackage{hypcap}
\usepackage{multirow}

\hypersetup{ 
colorlinks=true,
linkcolor=blue,
citecolor=blue
}

\begin{document}

\author{Brett Leedahl}
\affiliation{Department of Physics and Engineering Physics, University of Saskatchewan, 116 Science Place, Saskatoon, Saskatchewan S7N 5E2, Canada}
\email{brett.leedahl@usask.ca}

\author{Tristan de Boer}
\affiliation{Department of Physics and Engineering Physics, University of Saskatchewan, 116 Science Place, Saskatoon, Saskatchewan S7N 5E2, Canada}

\author{Xiaotao Yuan}
\affiliation{Beijing National Laboratory for Molecular Sciences and State Key Laboratory of Rare Earth Materials Chemistry and Applications College of Chemistry and Molecular Engineering. Peking University, Beijing 100871, China}

\author{Alexander Moewes}
\affiliation{Department of Physics and Engineering Physics, University of Saskatchewan, 116 Science Place, Saskatoon, Saskatchewan S7N 5E2, Canada}

\title{Oxygen Vacancy Induced Structural Distortions in Black Titania - A Unique Approach using Soft X-Ray EXAFS at the O K-Edge}

\begin{abstract}
Unknown changes in the crystalline order of regular TiO$_2$ result in the formation of black titania, which has garnered significant interest as a photocatalytic material due to the accompanying electronic changes. Herein, we determine the nature of the lattice distortion caused by an oxygen vacancy that in turn results in the formation of mid-band gap states found in previous studies of black titania. We introduce an innovative technique using a state-of-the-art silicon drift detector, which can be used in conjunction with extended x-ray absorption fine structure (EXAFS) to measure bulk interatomic distances. We illustrate how the energy dispersive nature of such a detector can allow us an unimpeded signal, indefinitely in energy space, thereby sidestepping the hurdles of more conventional EXAFS, which is often impeded by other absorption edges.
\end{abstract}

\maketitle 

\section{Introduction}
Structural information regarding the transformation that takes place in regular ``white'' titanium dioxide as it transitions to black titania is still not fully understood, despite significant interest.\cite{Coh2017} Although we know these crystallographic changes are key in the desirable electronic properties that arise, how the former causes changes the latter is not entirely clear.\cite{Liu2013} However, it has been shown that new XRD peaks appear as the white to black transformation occurs, undoubtedly indicating some sort of structural distortion.\cite{Lu2014} In addition to this, while the surface-disordered crystalline state of black titania has been documented via high resolution electron microscopy,\cite{Chen2013,Zhou2014,Yim2010} the bulk character is far less understood.\cite{Ataei2017} As such, the exact nature of the distortion present has been difficult to determine on account of the complexity and inhomogeneity of the crystal. 

Extended x-ray absorption fine structure (EXAFS) is the study of the oscillatory nature of the absorption coefficient as a function of incident photon energy at energies just beyond that of an absorption edge. These oscillations arise from the local atomic structure of a material, and are most frequently used to determine bond distances and vibrational properties of a given element in a material.\cite{Teo1986} However, despite being continuously refined ever since its discovery in the early 1970s,\cite{Stern1974} it still has yet to find commonplace use in the soft x-ray energy regime ($\approx$50-1500 eV), despite the technique having contributed to many significant advances.\cite{Popmintchev2018,Hu2013,Zhang2018} Only a small subset of examples exist in the publication record (all performed by surface sensitive total electron yield, or transmission mode in which extremely thin samples are required to pass soft x-rays through), and nothing in the last two decades since the advent of high resolution silicon drift detectors (SDDs).\cite{Stohr1978,Stohr1979,Stohr1979-2,Troger1993,Dagg1993,Amemiya1999,Yang1987,Kao1989,Zubavichus2004}

There are a few reasons for this. (1) There is an abundant number of absorption edges in this energy range. Every element heavier than helium contains electrons with binding energies in this range. This is generally considered a limitation and large problem for EXAFS experiments, as one requires a range of several hundred ($> 400$) electron-volts of unperturbed absorption data from a single edge to extract meaningful data. (2) Many soft x-ray beamlines at synchrotron facilities do not have a monochromator that can, in a single scan, slew a necessarily large energy range without some combination of changing the mirror and grating of that monochromator. (3) The low probability of fluorescence emission in low-Z elements implies that extremely long count times are required to obtain sufficiently noise-free spectra.

However, using a silicon drift detector, one can achieve very high fluorescence count rates at sufficiently high resolution in the soft x-ray regime.\cite{Lechner2001,Lechner2004} And because silicon drift detectors are energy dispersive, the intrusion of absorption edges impeding our energy range is inconsequential, because we are able to filter out emission photons from other elements and observe only the fluorescence photons from the element of interest. If, however, the sample contains a large enough atomic percentage of some absorption edge energetically above our measured edge (here O), the XAS signal will contain an inverted image of that element as a new decay pathway emerges; fortunately, steps can be taken to disregard this as well. Furthermore, since we are observing emitted photons (as opposed to electrons) in the soft x-ray regime, the technique is bulk sensitive, probing to $\approx$100 nm in depth.

\begin{figure}
\begin{center}
\includegraphics[width=3.375in]{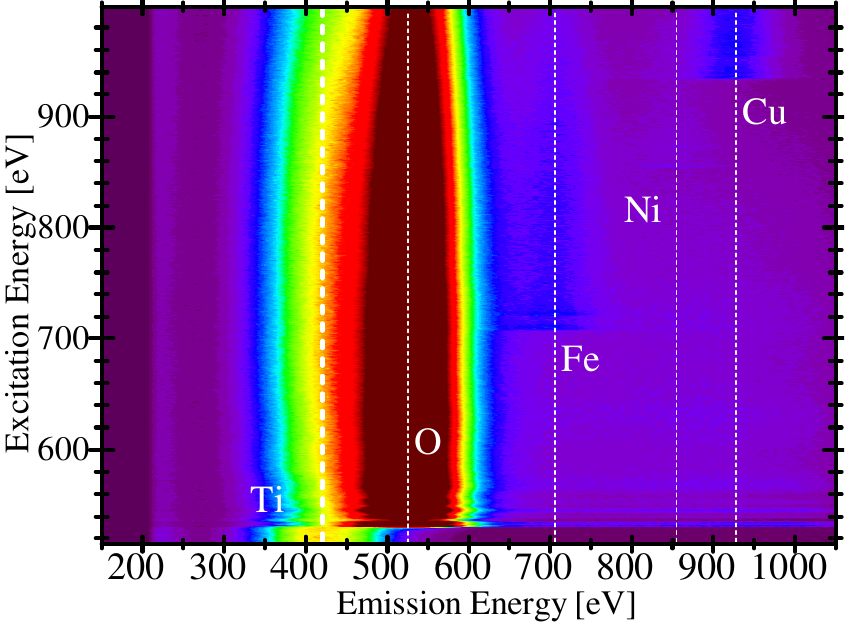}
\caption{Color map of the TiO$_2$ anatase spectrum. Each horizontal slice corresponds to a single emission spectrum recorded by the SDD. The Ti emission is clearly visible pre-O $K$-edge. Because of this overlap with the O $K$-edge emission, one must separate the $\approx$100 eV wide emission lines by a fitting procedure to isolate the oxygen emission photons. Note the anomalous emission photons at the Fe, Ni, and Cu $L_{2,3}$ edges at 707, 858, and 933 eV, respectively. These are unwanted and are neglected from the data analysis using the procedure described in the text.}
\label{fig:map}
\end{center}
\end{figure}


\section{Experiment and Calculation Methods}
All absorption spectra were measured at the Resonant Elastic and Inelastic X-Ray Scattering (REIXS) beamline at the Canadian Light Source. The spectra were recorded in fluorescence yield mode at room temperature using an SDD, such that the emission lines from the various elements could be distinguished on an energy scale. The total data set for a spectrum can be visualized in a two dimensional color map plot as shown in Figure \ref{fig:map}.  The energy resolution in this energy range is $\approx$100 eV, and therefore overlapping emission lines need to be dissociated.\cite{Tolhurst2017} 



The Fe, Ni, and Cu $L_{2,3}$ emission photons that are observable in Figure \ref{fig:map} arise from scattered photons in the sample chamber interacting with the chamber itself. The ultra high vacuum chamber is made of stainless steel, and the arm that holds the sample is largely made of copper. The high intensity synchrotron light that impinges on the sample is scattered in all directions, but the SDD detector will detect only a minute solid angle of the overall emitted and scattered photons. However, these stray photons will excite all constituent materials of the chamber as well. Since an SDD is not an angle resolved detector, it records these secondary photons as well. 

\begin{figure}
\begin{center}
\includegraphics[width=3.375in]{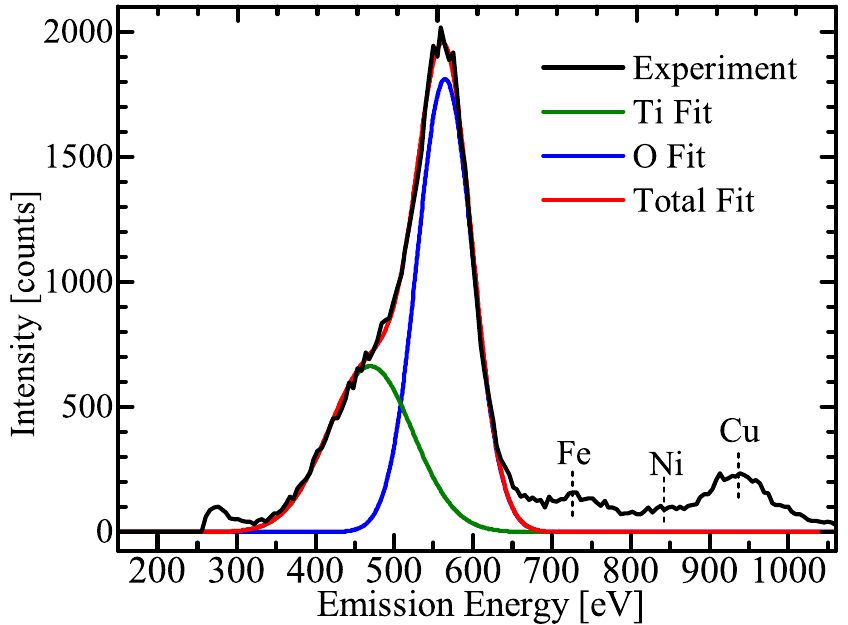}
\caption{The experimental spectrum shown here corresponds to a single horizontal slice in Figure \ref{fig:map} taken at 975 eV excitation energy. In order to resolve them we used a fitting algorithm that constrains both the peak energy and FWHM for the Ti and O emission lines. The integrated intensity of the O peak corresponds to a single data point in the partial fluorescence yield XAS spectra shown in the Supplemental Material. Note that unwanted fluorescence from Fe, Ni, and Cu are easily neglected using this method.}
\label{fig:fit}
\end{center}
\end{figure}

While these secondary emissions are generally undesirable, for our present purposes it illuminates our ability to disregard unwanted photons, whether in our sample or otherwise. We can show that in-sample elements such as Ti can be excluded from the analysis, as well as out-of-sample materials that cannot be avoided such as Fe, Ni, and Cu. 

The separation of these emission lines can be accomplished with a constrained Gaussian fitting algorithm, wherein the relevant peaks are each fit by a Gaussian curve, and the integrated Gaussian corresponds to the number of photons emitted by a given element for a given incident energy. A Gaussian curve contains three relevant parameters: its center, width (FWHM), and intensity. For the present analysis the Ti and O FWHMs and centers were constrained to remain in a narrow range as to not allow the fitting to stray from reasonable values.



The post-experiment analysis was performed with EXAFSPAK software \cite{George1995} wherein a spline function was subtracted from the raw spectra (see Supplemental Material for raw experimental spectra) to obtain the EXAFS signal (red lines in Figure \ref{fig:exafs}), which has the energy axis converted to wavenumber $k$. The background spline subtraction was optimized to reduce the spurious peaks that appear in the Fourier transform at low $R$ values (less than the nearest neighbour bond distances). The displayed EXAFS signals are also weighted by their $k^2$ value, where $k^2=2m_eE/\hbar^2$ and $m_e$, $E$, and $\hbar$ are the electron mass, incident photon energy, and reduced Planck's constant, respectively. McKale table theoretical phase-shift and amplitude functions of the absorber-backscatter interaction were used for rutile, while full FEFF calculations for the backscattering amplitudes and phases were run to produce the models for anatase and black titania.\cite{McKale1988} To achieve the fits to experiment shown in Figure \ref{fig:exafs}(c-f), only the interatomic distances and Debye-Waller factors were allowed to vary. Moreover, the Debye-Waller factors---which is a measure of the mean squared deviation of an atom from its equilibrium position averaged over all atoms---was only allowed to vary within chemically reasonable values centered around 0.005 {\AA}$^2$.

The synthesis of our black titania sample has been extensively discussed in previous publications.\cite{Liu2016,Yuan2016,Wang2013} Oxygen deficient black titania was prepared by an Al-reduction method in which amorphous TiO$_2$ (prepared by hydrolysis of TiCl$_4$) and Al powder were placed separately in a two zone tube furnace and then evacuated to a base pressure lower than 0.5 Pa. After that, the aluminum was heated beyond its melting point to 800$^\circ$C, and the TiO$_2$ compartment was heated to 500$^\circ$C for 4 hours. Thermodynamically, the reaction driving force enables aluminum oxidation and titania reduction (TiO$_2$ + Al $\rightarrow$ TiO$_{2-x}$+AlO$_x$).



\section{Results and Discussion}
\subsection{Rutile and Anatase TiO$_2$}
To demonstrate the reliability of this method, the well-known rutile and anatase TiO$_2$ powders were measured and compared to the known structures. Generally one wishes to obtain nearest and perhaps second nearest neighbor distances. However, we show that using the high count rates of modern SDD detectors, we can obtain interatomic distances well beyond this threshold in a reasonable amount of time. 

Given the relationship that the resolution in real space is $\Delta R=\pi/2\Delta k$,\cite{Calvin2013} we obtained spectra of significant quality to achieve resolutions of 0.19 {\AA}. One should be careful to distinguish these values from the accuracy with which bond lengths can be determined, which is generally on the order of a few hundredths of an angstrom. The real space resolution values only tell us that peaks in the Fourier transform (interatomic distances) that are not separated by more than this distance cannot be distinguished from one another. On the other hand, we propose that our experimental error is $\pm$0.02 {\AA} for O--Ti distances. This can readily be defended on account of the EXAFS analysis in which several dozen combinations of background subtractions, $k$-ranges, and smoothing parameters (or lack thereof) were used to produce the final Fourier transform. The result was that the peaks in the final Fourier transform analysis would only vary by $\pm$0.02 {\AA} using any reasonable combination parameters. Random errors due to the noise in the experiment result in errors roughly an order of magnitude less than this, and are thus negligible.

\begin{figure}
\begin{center}
\includegraphics[width=3.375in]{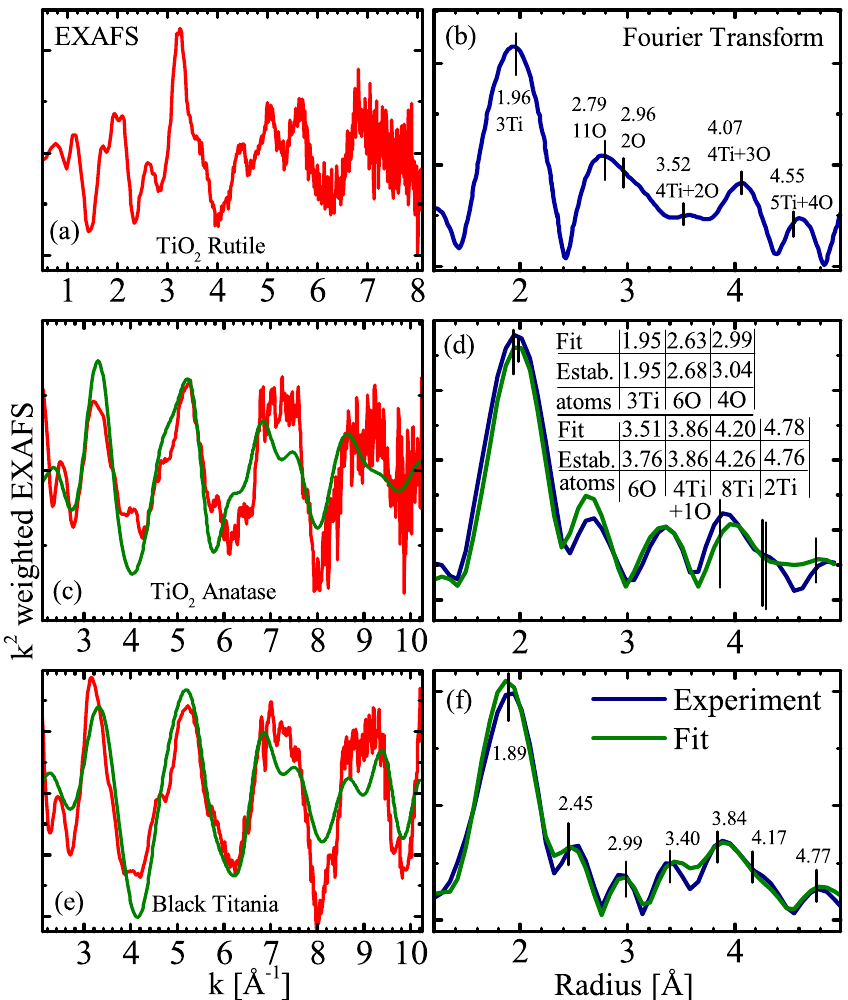}
\caption{(a)(c)(e) EXAFS signal derived from XAS after subtracting a spline curve from the raw spectra. (b)(d)(f) Fourier transforms of EXAFS signals. For rutile, vertical lines represent where one would expect peaks to appear given the well established structure. For anatase, vertical lines were placed at the known Ti-O distances, with the length of line representing Ti degeneracy at that distance. The fit to the anatase experimental EXAFS was obtained using the interatomic distances shown in panel (d), which compare favorably to the established values (which are weighted averages of the nominal values shown in Table \ref{tbl:bondlengths}). For the unknown black titania structure in (f), the vertical lines are located at the interatomic distances that were used to obtain to fits. 
}
\label{fig:exafs}
\end{center}
\end{figure}

Figure \ref{fig:exafs} shows the $k^2$ weighted EXAFS, and the resulting Fourier transforms. The vertical lines in the Figure \ref{fig:exafs}(b) show where the \emph{peaks ought to be located in the nominal crystal structure of rutile}. Given that some coordination spheres will contain both Ti and O atoms at nearly the same distance from the central O atom, and noting our finite energy resolution ($\Delta R$), we would expect some peaks in the Fourier transform to contain a weighted average of the atoms that are at similar distances from the central oxygen atom. This is reflected by the vertical lines for rutile, which makes it visually easy to identify our experimental peak locations in relation to the nominal crystal of rutile, which is well known. The agreement is easily accurate to within the small discrepancies concerning these distances that exist in the literature, which vary by about 2\%.\cite{Meagher1979,Smyth1995,Vasquez2016} 

In both rutile and anatase TiO$_2$ the nearest neighbor Ti atoms can easily be determined to within 0.01 {\AA} of the nominal value. The accuracy in determining this nearest neighbour distance is often all that one expects from EXAFS experiments. However, by comparison of the actual peak locations from our experiment to the nominal values indicated by the vertical lines in Figure \ref{fig:exafs}(b) and the comparison of modelled and the established values in panel (d), it can be seen that further coordination shells can also be quite reliably determined with this method. Furthermore, the experiments were repeated and are entirely reproducible, indicating the consistency and reliability of this method.

\subsection{Black Titania}
While x-ray diffraction has shown there is indeed some disorder induced lattice strain in antase TiO$_2$ when transitioning to black titania, quantifying it has eluded the scientific world.\cite{Wang2013} We have used a model containing oxygen vacancies,\cite{Maqbool2017,Janotti2010} which has many times in the past been shown to be an energetically favorable defect in TiO$_2$,\cite{Kowalski2010,Pan2013} and is therefore probable to form in the case of black titania.\cite{Phattalung2006} We offer experimental evidence to supplement the theoretical work of this claim. We propose that the relaxation of atomic positions in the vicinity of this vacancy will cause deviations of the nominal bond lengths of anatase. Predictably, this occurs in a quite complex fashion; as one bond angle and/or distance changes, it has the propensity to alter those in its vicinity, and so on to the next nearest neighbours throughout the crystal.

\begin{figure}
\begin{center}
\includegraphics[width=3.375in]{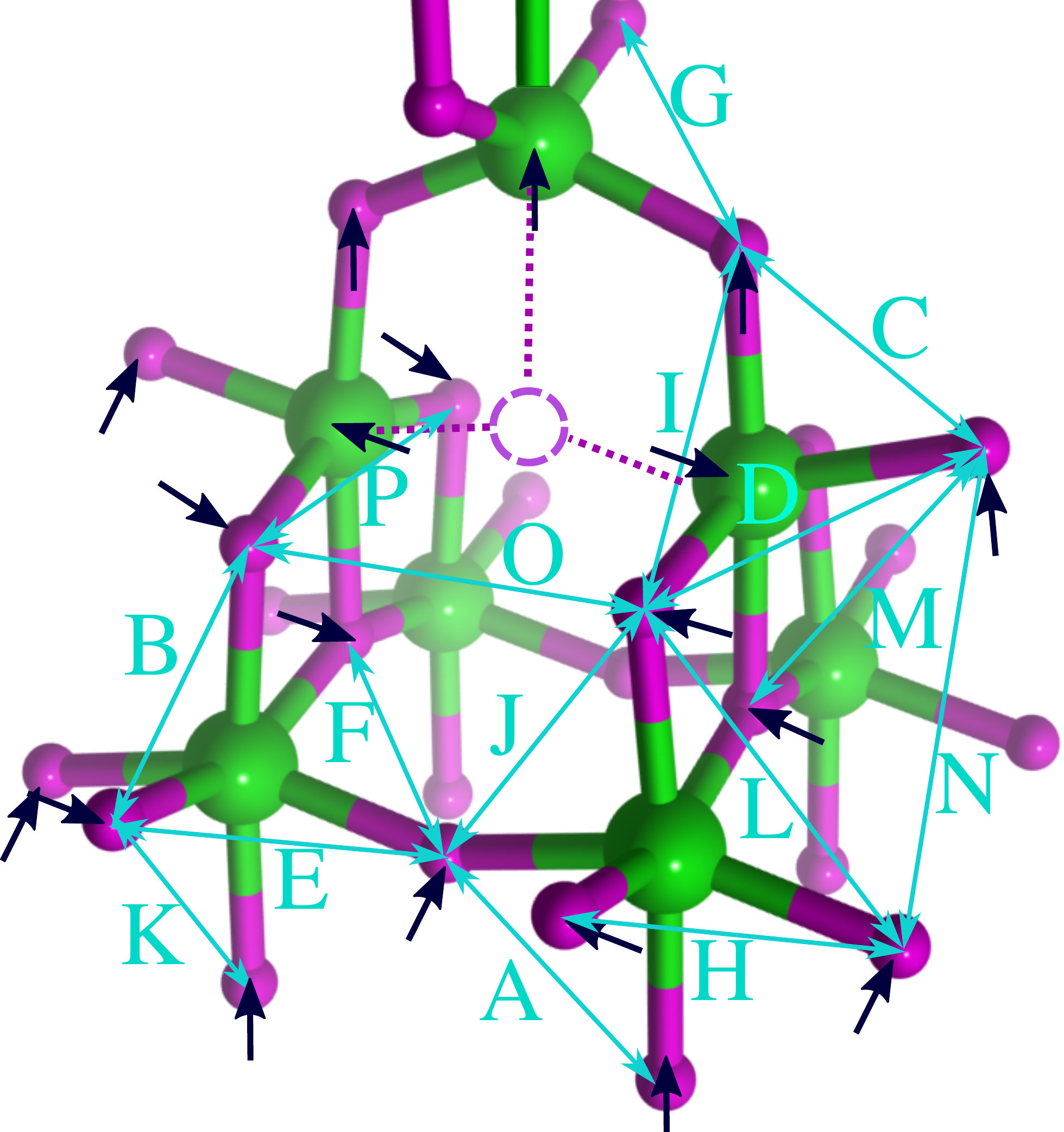}
\caption{Proposed distorted anatase structure that results in black titania and its electronic properties. Arrows indicate the direction of  the spatial relaxation of the atoms from their nominal positions. The labeled interatomic distances correspond to those listed in Table \ref{tbl:bondlengths}.}
\label{fig:blackstructure}
\end{center}
\end{figure}

We have formulated a defect that is well supported by the experimental data. This defect, which includes an oxygen vacancy as well as the shift of the nearby atoms, is shown in Figure \ref{fig:blackstructure}. And as there are no inequivalent oxygen sites in anatase, all oxygen atoms are equivalent, and so removing any oxygen atom from the nominal anatase structure is equivalent. The advantage of using EXAFS in comparison to other techniques often used as evidence of oxygen vacancies such as electron paramagnetic resonance,\cite{Song2017} is that we can determine experimentally the actual crystal distortion on the scale of hundredths of nanometers.
 
Black titania is generally synthesized under a hydrogen rich environment and is known to have an enhanced hydrogen mobility thereafter.\cite{Chen2013,Yang2013,Wang2013} This is likely to result in bulk oxygen vacancies  which give rise to itinerant H$_2$ molecules and the appearance of OH bonds in the bulk, both of which play a crucial role in forming the mid-band gap states in black titania that make it such an attractive material. These weakly bound hydrogen atoms are able to easily diffuse throughout the crystal via the lattice distortions introduced here. The proclivity towards oxygen vacancy formation in TiO$_2$, combined with the hydrogenation process allow a subtle alteration to crystal structure, which in turn leads to a drastic alteration of the electronic properties.

\begin{table}
\centering
\begin{tabular}{| l | c |  c | c | c |}
    \hline
 & Nominal & Degeneracy & Distorted & Multiplicity\\ \hline
 & 1.95 & 3 Ti & 1.92(avg.) & 36 Ti\\ \hline
A &   & \multirow{3}{*}{2 O} & 2.45 & 8 O \\ 
B & 2.46 & & 2.43 & 8 O \\ 
C &   &  & 2.50 & 2 O\\ \hline
D &   & \multirow{5}{*}{4 O} & 2.95 & 4 O \\ 
E &   & & 2.75 & 8 O \\ 
F & 2.79 & & 2.60 & 4 O \\
G &   &  & 2.76 & 2 O \\ 
H &   &  & 2.91 & 4 O \\ \hline
I &   & \multirow{5}{*}{4 O}& 3.21 & 4 O \\ 
J &   & &  2.73 & 4 O\\
K & 3.04 & & 2.88 & 8 O \\ 
L &   & &  2.99 & 4 O \\
M &   & & 3.13 & 2 O \\ \hline
N & 3.71 & 2 O & 3.62 & 4 O \\  \hline 
O &   & \multirow{4}{*}{4 O} & 3.48 &2 O\\
P & 3.78 & & 3.78 & 20 O \\
 &   & & 4.09 & 4 O \\ \hline
 & 3.86 & 4 Ti & 3.84(avg.) & 16 Ti \\  \hline
& 4.26 & 8 TI & 4.20(avg.) & 46 Ti \\ \hline
& 4.76 & 2 Ti & 4.76(avg.) & 10 Ti \\ \hline
\end{tabular}
\caption{Interatomic distances (in {\AA}) are as labeled in Fig \ref{fig:blackstructure}, \emph{distorted} distances are taken from the theoretically proposed structure of Figure \ref{fig:blackstructure}. The nominal structure corresponds to that of anatase, black titania's parent structure. \emph{Degeneracy} refers to the number of atoms in the coordination sphere for every oxygen atom in the nominal structure. When a vacancy is introduced, the degeneracy of the distances in the anatase structure is broken and are split into several different distances in the distorted structure, which we call multiplicity. \emph{Multiplicity} refers to the number of oxygen-oxygen or oxygen-titanium distances \emph{per oxygen defect site}. The letters correspond to those shown in Figure \ref{fig:blackstructure}.}
\label{tbl:bondlengths}
\end{table}

Our experimental EXAFS spectrum for black titania, along with its Fourier transform is displayed in Figure \ref{fig:exafs}(e-f). The vertical lines in panel (f) are now experimentally determined bond lengths found via the best fit algorithm described above. As a distortion is introduced to the lattice, the degeneracy of oxygen-oxygen distances in anatase is broken, and instead of six coordination spheres between 2 {\AA} and 4 {\AA} in anatase, many more will occur in the vicinity of a vacancy (see Table \ref{tbl:bondlengths}). In the resulting Fourier transform for black titania, this will lead to smearing of peaks and the appearance of peaks at roughly the average of several of these combined distances. While this degeneracy is also broken in the case of Ti-O distances, it is much more manageable since the resulting Ti-O distances do not stray significantly from their central value (i.e. all the previously 4.26 {\AA} bonds end up closely bunched around 4.17 {\AA}). This, in addition to the fact that the Ti backscattering amplitude is much larger than O backscattering as a consequence of Ti's larger atomic size, implies that the Ti-O derived distances are more reliable than the O-O distances. For these reasons, the Ti-O distances were used as the basis for the proposed structural defect found in Figure \ref{fig:blackstructure}. That is, our proposed black titania distortion was found such that it agrees with the interatomic Ti-O distances found via experiment and fitting. Table \ref{tbl:bondlengths} lists the bond lengths in nominal anatase as well as the degeneracy and type of atom that exists for each coordination sphere, and the equivalent information in the vicinity of a vacancy for our proposed structure. The labeled bonds A to P correspond to those illustrated in Figure \ref{fig:blackstructure}, the unlabeled ones were not shown in the figure as they would significantly obstruct clarity.

The first point to note is the significant reduction that occurs in the bond length to the nearest Ti neighbours: two Ti atoms at 1.94 {\AA} and one at 1.98 {\AA} now become resolved at 1.89 {\AA} in Figure \ref{fig:exafs}(f). The next Ti coordination sphere for anatase is at 3.86 {\AA}, which we found at 3.84 {\AA} in black titania, which is effectively identical within experimental error, which is $\pm$0.02 {\AA} for O--Ti distances. However, a noticeable reduction happens for the next Ti coordination sphere at 4.26 {\AA} in anatase, but is found at 4.17 {\AA} in our experiment. Lastly, the 4.78 {\AA} Ti-O distance remains unchanged in the transition to black titania. While the trend is not consistent, this is actually encouraging from the perspective of searching for potential defect structures, as it vastly limits the number of possibilities. 

Hundreds of millions of defect structures were tested by combing through the symmetry-reduced parameter space and individually comparing the distances in the model defect structures with the experimental results and choosing the closest fit, wherein a single vacancy was introduced and the atoms within 5 {\AA} were allowed to shift around it. That is, the interatomic distances that resulted from simulations were compared to those found via experiment and fitting. To determine the amount of movement allowed by the lattice, DFT calculated relaxed structures using WIEN2k \cite{Schwarz2002} with a PBEsol functional \cite{Perdew2008} (with a 3\% oxygen vacancy density) showed that nearest neighbour Ti-O bond lengths changed by up to 0.125 {\AA} upon the introduction of a vacancy, while farther Ti-O distances were changed by up to 0.25 {\AA}. Therefore, the candidate structures were generated by allowing the atoms around the defect to move randomly within a reasonable range of less than 0.25 {\AA} such that the symmetry around the oxygen vacancy remained intact. The conclusion is that while DFT relaxed structures alone were inconclusive, the only possible solution given the combined knowledge of DFT and generated candidate structures is the scenario is that shown in Figure \ref{fig:blackstructure}. 

The qualitative result is that Ti atoms, on account of their now dangling bonds, shift slightly away from the vacancy in order to strengthen their bond with the rest of the lattice, while the nearby O atoms shift slightly inwards toward the vacancy to fill the empty space. This distortion \emph{uniquely} maintains the above stated trends displayed by the observed shifts (and non-shifts) of Ti-O interatomic distances in the experimental data, as well as takes into account the information gained through DFT calculations. 


It is necessary to note that we should only expect near perfect agreement if it were the case that this distortion were repeated exactly throughout the entire crystal. However, in real world black titania the defect is not perfectly ordered throughout. And since the EXAFS results are a bulk average of all O atoms in the material, we should expect some influence from non-distorted sites. However, given that our result is very accurate---with the model displaying the correct trends given by the experimental data---it is surely a true representation of the crystal structure of black titania.

\section{Conclusion}
Herein, we have shown conclusively that bulk soft x-ray EXAFS at the O $K$-edge is a valuable tool for systematically determining subtle structural distortions. We have presented very strong evidence that structural changes resulting from vacancies in a crystal can be directly measured. These changes to a parent material often underpin the emergence of novel electronic properties, and are often of the utmost importance to understand. This technique should not be overlooked, and has become feasible with the advent of modern silicon drift detectors, for which one can sidestep some of the difficulties inherent to EXAFS experiments.

\section{Acknowledgements}
This work was supported by the Natural Sciences and Engineering Research Council of Canada (NSERC) and the Canada Research Chairs program. Measurements were performed at the Canadian Light Source (supported by NSERC and the University of Saskatchewan).

\end{document}